\documentclass{emulateapj}
\usepackage{apjfonts}

\slugcomment{June 16, 2008; revised August 5, 2008; accepted August
  11, 2008}

\shorttitle{RADIATION FROM PRIMARY ELECTRONS IN SUPERNOVAE}
\shortauthors{ANDO \& M{\'E}SZ{\'A}ROS}

\begin{document}

\title{Broadband Radiation from Primary Electrons in Very Energetic
  Supernovae}
\author{Shin'ichiro Ando}
\affil{California Institute of Technology, Mail Code 130-33, Pasadena,
  CA 91125}
\email{ando@tapir.caltech.edu}
\and
\author{Peter M{\'e}sz{\'a}ros}
\affil{Department of Astronomy and Astrophysics, Pennsylvania State
  University, University Park, PA 16802}
\affil{Department of Physics, Pennsylvania State University,
  University Park, PA 16802}
\affil{Center for Particle Astrophysics, Pennsylvania State
  University, University Park, PA 16802}
\email{nnp@astro.psu.edu}

\begin{abstract}
A class of very energetic supernovae (hypernovae) is associated
with long gamma-ray bursts, in particular with a less energetic but
more frequent population of gamma-ray bursts. Hypernovae also
appear to be associated with mildly relativistic jets or outflows,
even in the absence of gamma-ray bursts. Here we consider radiation 
from charged particles accelerated in such mildly relativistic outflows 
with kinetic energies of $\sim$10$^{50}$ erg. The radiation processes 
of the primarily accelerated electrons considered are synchrotron 
radiation and inverse-Compton scattering of synchrotron photons
(synchrotron self-Compton; SSC) and of supernova photons (external 
inverse-Compton; EIC). In the soft X-ray regime, both the SSC and 
EIC flux can be the dominant component, but due to their very different
spectral shapes it should be easy to distinguish between them.
When the fraction of the kinetic energy going into the electrons
($\epsilon_e$) is large, the SSC is expected to be important; otherwise 
the EIC will dominate. The EIC flux is quite high, almost independently of
$\epsilon_e$, providing a good target for X-ray telescopes such as
{\it XMM-Newton} and {\it Chandra}.  In the GeV gamma-ray regime, the 
EIC would be the dominant radiation process and the {\it Gamma-ray Large
Area Space Telescope (GLAST)} should be able to probe the value of
$\epsilon_e$,
the spectrum of the electrons, and their maximum acceleration energy.
Accelerated protons also lead to photon radiation through the secondary 
electrons produced by the photopion and photopair processes. We find that
over a significant range of parameters the proton component is 
generally less prominent than the primary electron component. We discuss
the prospects for the detection of the X-ray and GeV signatures of 
the mildly relativistic outflow of hypernovae. 
\end{abstract}


%%%%%%%%%%%%%%%%%%%%%%%%%%%%%%%%%%%%%%%%%%%%%%%%%%%%%%%%%%%%%%%%%%%%
\section{Introduction}
\label{sec:Introduction}

While it is recognized that long gamma-ray bursts (GRBs) are
associated with very energetic supernovae, sometime referred to as
hypernovae, our understanding of the physics that drives and connects
these events is far from being established.  Observationally, long GRBs 
are more complicated than were thought before. For example, they appear 
to have subgroup that includes GRB 980425 \citep{Galama1998}, GRB 031203
\citep{Malesani2004,Soderberg2004} and GRB 060218
\citep{Campana2006,Cobb2006,Pian2006}.
These GRBs occurred relatively nearby and their energies radiated by
prompt gamma rays were significantly smaller than the other
long GRBs, and they were associated with well-studied hypernovae (SN
1998bw, SN 2003lw and  SN 2006aj, respectively). Radio observations 
of these events \citep{Kulkarni1998,Soderberg2006} suggest the presence 
of mildly relativistic ejecta, which is a different component from 
the usual nonrelativistic component of the supernova explosion.
Their rate of occurrence may be an order of magnitude higher than that 
of the more energetic GRBs \citep{Liang2007,Soderberg2006}.
See also, e.g.,
\citet*{Liang2006,Murase2006,Toma2007,Waxman2007,Gupta2007} for other
followup studies.
Very recently, a mildly  relativistic outflow component has also been
inferred in a supernova of type Ibc unassociated with a GRB, SN 2008D
\citep{Soderberg2008}.

In this paper, we investigate the broadband radiation from the
mildly relativistic ejecta associated with hypernovae, focusing 
especially on the high-energy photon emission in the X-ray and 
gamma-ray ranges.  We mainly study the radiation from relativistic 
electrons which are primarily accelerated in shocks, but we also 
consider the radiation from secondary electrons which are generated 
by interactions involving accelerated protons. 
The latter were extensively studied by \citet{Asano2008} in connection
with the origin of Galactic cosmic rays \citep{Wang2007,Budnik2008}.
Our treatment of the proton component includes also the previously
neglected effect of photopair production. We show that the primary electron 
component photon signature generally dominates over the proton component,
for a wide range of energies.  In particular the inverse-Compton 
scattering of hypernova photons due to the primary electrons appears the 
most promising channel for detection in the X-ray and gamma-ray ranges.

The paper is organized as follows.
In \S~\ref{sec:Spectrum of Primary Electrons}, we introduce 
the relevant supernova parameters and the spectrum of the primary
accelerated electrons.
We present our main results on the radiation from the primary 
electrons in \S~\ref{sec:Radiation from Primary Electrons}, 
and discuss its dependence on the parameters.
Section~\ref{sec:Radiation from secondary electrons and proton
acceleration} is devoted to a discussion of the photon radiation 
of a proton origin and its comparison to the electron component.
We summarize our conclusions in \S~\ref{sec:Conclusions}.

\section{Spectrum of Primary Electrons}
\label{sec:Spectrum of Primary Electrons}

\subsection{Supernova dynamics}
\label{sub:Supernova dynamics}

As suggested by \citet{Soderberg2006}, there is an empirical relation
between the kinetic energy ($E_K$) and the velocity ($\beta c$) of the 
hypernova ejecta, parameterized as $E_K = 10^{52} (\Gamma \beta / 0.1)^{-2}$, 
where $\Gamma = (1 - \beta^2)^{-1/2}$.  The low velocity component with 
$\Gamma \beta \simeq 0.1$ corresponds to a very energetic supernova, 
while the high velocity component with $\Gamma \beta \simeq 1$ corresponds
to the mildly relativistic ejecta.  It is assumed that the energy of the
ejecta components is dissipated when they start to decelerate, going into 
relativistic particles.

The deceleration occurs as a result of interaction with the surrounding
matter, which we assume is provided by the mass loss of the massive 
progenitor (Wolf-Rayet type) star.  We adopt a fiduciary mass-loss rate 
of $\dot M = 10^{-5} M_{\sun}$ and a wind velocity $v_w = 10^3$ km s$^{-1}$.
The density profile is then $\rho(r) = (\dot M / 4 \pi v_w) r^{-2} 
= 5 \times 10^{11} r^{-2}$ g cm$^{-1}$.  In this case, the kinetic energy 
($E_K = 10^{50}$ erg) of the mildly relativistic ejecta ($\Gamma \beta = 1$) 
starts to be dissipated when it reaches at radius $r = R \simeq
10^{16}$ cm.
The corresponding dynamical time scale is $t_{\rm dyn} = R/ c \Gamma
\beta = 3 \times 10^5$ s, a few days.  Around that time, the hypernova 
luminosity is still very large.  On the other hand, if we consider 
sub-relativistic but more energetic ejecta, the dynamical time scale is
significantly  larger, by which time the hypernova luminosity has
decreased, a situation that is not interesting for the present
purposes.

The hypernova luminosity around a few days after the explosion is roughly
$L_{\rm SN} = 10^{43}$ erg s$^{-1}$, and we assume the spectrum is
black body with a temperature of $\sim$1 eV; a typical photon energy is
$\varepsilon_\gamma = 2.7$ eV.
The photon number density at the dissipation radius $R$ is then given by
$n_\gamma = L_{\rm SN} / \pi c R^2 \varepsilon_\gamma = 4 \times
10^{11}$ cm$^{-3}$.  We note that the parameter values above are very 
similar to those in \citet{Wang2007} and \citet{Asano2008}.

\subsection{Acceleration and cooling of primary electrons}
\label{sub:Acceleration and cooling of primary electrons}

When dissipation starts, we assume a fraction $\epsilon_e$ of the
kinetic energy goes into primary electrons accelerated to relativistic
speed in the shock, and their initial spectrum is a power law with
index $-p$.
A typical Lorentz factor (in the ejecta frame) of the primary
electrons is then given by $\gamma_m = \epsilon_e (m_p / m_e) \Gamma =
260 (\epsilon_e/0.1)$.
In addition, we assume a fraction $\epsilon_B$ of $E_K$ goes into
magnetic fields, yielding $B = [8 \pi \epsilon_B \rho(R) c^2]^{1/2} =
1 (\epsilon_B/10^{-2})^{1/2}$ G.
We shall study the dependence of results on these parameters in the
following, but unless stated, we adopt $\epsilon_e = 0.1$, $\epsilon_B
= 10^{-2}$, and $p = 2.5$ as fiducial values, according to the analogy
to GRB emission.

After acceleration, these electrons immediately lose their energies
through radiation unless the cooling time scale $t_{\rm cool}$ is
longer than the dynamical time scale $t_{\rm dyn}$.
Relevant radiation mechanisms include synchrotron radiation due to
magnetic field, inverse-Compton scatterings off hypernova photons
(external inverse-Compton; EIC) and synchrotron photons (synchrotron
self-Compton; SSC).
The ratio of cooling time scales of EIC and synchrotron processes is
given by $t_{\rm EIC} / t_{\rm syn} = U_B / \varepsilon_\gamma
n_\gamma \approx 0.05 (\epsilon_B / 10^{-2})$.
In addition, unless $\epsilon_e$ is much larger than $\epsilon_B$, the
SSC cooling time scale is at most comparable to that of synchrotron
radiation. Thus, among those three, EIC is the most efficient
mechanism for electron energy losses, and we have $t_{\rm cool}
\approx t_{\rm EIC} < t_{\rm dyn}$ for $\gamma_e > \gamma_c \simeq
70$, where $\gamma_e$ represents the electron Lorentz factor in the
ejecta frame.

When the electrons are in the ``fast-cooling'' regime, defined as the
case when $\gamma_c < \gamma_m$ (or equivalently $\epsilon_e > 0.03$ in 
the current context), the electron spectrum is given by
\begin{equation}
\frac{dN_e}{d\gamma_e} = \frac{N_e}{\gamma_c}\times \left\{
\begin{array}{ll}
\left(\frac{\gamma_e}{\gamma_c}\right)^{-2} & [\gamma_c < \gamma_e <
  \gamma_m],\\
\left(\frac{\gamma_m}{\gamma_c}\right)^{-2} \left(\frac{\gamma_e}
  {\gamma_m}\right)^{-p-1} & [\gamma_e > \gamma_m],
\end{array}
\right. 
\label{eq:Ne fast cooling}
\end{equation}
where the normalization is set so that we have correct number of
electrons $N_e = E_K / \Gamma m_p c^2 = 5 \times 10^{52}$ after
integration \cite[e.g.,][]{Sari2001}.
On the other hand, in ``slow-cooling'' regime ($\gamma_c > \gamma_m$),
we have
\begin{equation}
\frac{dN_e}{d\gamma_e} = \frac{(p-1)N_e}{\gamma_m} \times \left\{
\begin{array}{ll}
\left(\frac{\gamma_e}{\gamma_m}\right)^{-p} & [\gamma_m < \gamma_e <
  \gamma_c],\\
\left(\frac{\gamma_c}{\gamma_m}\right)^{-p} \left(
  \frac{\gamma_e}{\gamma_c} \right)^{-p-1} & [\gamma_e > \gamma_c].
\end{array}
\right.
\label{eq:Ne slow cooling}
\end{equation}

\section{Radiation from Primary Electrons}
\label{sec:Radiation from Primary Electrons}

\subsection{Synchrotron radiation}
\label{sub:Synchrotron radiation}

The synchrotron mechanism gives the dominant contribution at radio
wavebands, and also provides seed photons for the SSC process.
The typical frequency and power from an electron with $\gamma_e$ is
\begin{eqnarray}
\nu_{\rm syn}(\gamma_e) & = & \frac{3 e B}{4 \pi m_e c} \gamma_e^2
\Gamma , \\
P_{\rm syn}(\gamma_e) & = & \frac{c \sigma_T}{6 \pi} \gamma_e^2 B^2
\Gamma^2 ,
\end{eqnarray}
where $\sigma_T$ is the Thomson cross section \citep{Rybicki1979}, and
we define $\nu_m \equiv \nu_{\rm syn} (\gamma_m)$ and $\nu_c \equiv
\nu_{\rm syn}(\gamma_c)$.

The flux of synchrotron photons $F_\nu^{\rm syn}$ is then given by
\begin{equation}
\frac{F_\nu^{\rm syn}}{F_{\nu,{\rm max}}^{\rm syn}} = \left\{
\begin{array}{ll}
\left(\frac{\nu}{\nu_c}\right)^{1/3} &
[\nu < \nu_c] ,\\
\left(\frac{\nu}{\nu_c}\right)^{-1/2} &
[\nu_c < \nu < \nu_m] ,\\
\left(\frac{\nu_m}{\nu_c}\right)^{-1/2}
\left(\frac{\nu}{\nu_m}\right)^{-p/2} & [\nu > \nu_m] ,\\
\end{array}
\right.
\label{eq:fnu fast cooling}
\end{equation}
for the fast-cooling phase, and
\begin{equation}
\frac{F_\nu^{\rm syn}}{F_{\nu,{\rm max}}^{\rm syn}} = \left\{
\begin{array}{ll}
\left(\frac{\nu}{\nu_m}\right)^{1/3} &
[\nu < \nu_m] ,\\
\left(\frac{\nu}{\nu_m}\right)^{-(p-1)/2}
& [\nu_m < \nu < \nu_c] ,\\
\left(\frac{\nu_c}{\nu_m}\right)^{-(p-1)/2}
\left(\frac{\nu}{\nu_c}\right)^{-p/2} & [\nu > \nu_c] ,\\
\end{array}
\right.
\label{eq:fnu slow cooling}
\end{equation}
for the slow-cooling phase, where $F_{\nu,{\rm max}}^{\rm syn} = N_e P_{\rm
syn}(\gamma_e) / 4\pi d^2 \nu_{\rm syn}(\gamma_e)$, and $d$ is distance
to the source \citep*{Sari1998}.
Here we neglected the effect of synchrotron self-absorption, which
however might become important at low-frequency radio bands.

We note that the radio flux at 10 GHz with this model and $d = 100$
Mpc is 1600 mJy for $\epsilon_e = 0.1$ and $\epsilon_B = 10^{-2}$
and 10 mJy for $\epsilon_e = 10^{-2}$ and $\epsilon_B = 10^{-4}$;
the self-absorption is irrelevant at this frequency.
For comparison, the radio fluxes around similar frequency at a peak
time were $0.5$ mJy for GRB 060218 \citep{Soderberg2006} and $50$ mJy
for GRB 980425 \citep{Kulkarni1998}.
Given that the radio flux strongly depends on both $\epsilon_e$ and
$\epsilon_B$, and that this model predicts a substantial flux, radio
observations would enable strong test to constrain these parameters or
others.

If the magnetic fields are highly inhomogeneous on very small spatial
scales, then the photon spectrum would be subject to another type of 
emission mechanism---jitter radiation \citep{Medvedev2000}.
% from relativistic electrons that gyrate
%in the magnetic field
In this case, the spectrum would be harder in the low frequency range,
$F_\nu \propto \nu$ below a jitter break frequency $\nu_{jm}$ (compare
this with the $\nu^{1/3}$ dependence of the synchrotron radiation).
Above $\nu_{jm}$, on the other hand, the spectrum is the same as that
of synchrotron.
The break frequency $\nu_{jm}$ is typically larger than the
synchrotron frequency $\min [\nu_c,\nu_m]$.
The difference in these radiation mechanisms could be tested from
observations. However,  in the following, we assume that the magnetic 
fields are quasi-homogeneous, so the synchrotron radiation is preferred.
This is because the shock waves we consider are close to the nonrelativistic
limit, where the generated magnetic fields would be more isotropic.
Such a situation is far from that of ultrarelativisitic GRBs
where the jitter radiation has been mainly considered.

\subsection{Inverse-Compton scattering}
\label{sub:Inverse-Compton scattering}

The SSC flux is straightforwardly obtained from the electron spectrum
(eqs.~[\ref{eq:Ne fast cooling}] and [\ref{eq:Ne slow cooling}]) and
the synchrotron flux (eqs.~[\ref{eq:fnu fast cooling}] and [\ref{eq:fnu
slow cooling}]) through the expression
\begin{equation}
F_\nu^{\rm SSC} = \frac{3\sigma_T}{4\pi R^2} \int d\gamma_e
\frac{dN_e}{d\gamma_e} \int_0^1 dx g(x) F_{\nu_s}^{\rm syn}
\left(\nu_s = \frac{\nu}{4\gamma_e^2 x}\right),
\end{equation}
where $g(x) = 1 + x - 2 x^2 + 2 x \ln x$ \citep{Rybicki1979,Sari2001}.
What makes the argument here simpler compared with the usual GRB SSC
model in the literature
\citep*[e.g.,][]{Sari2001,Zhang2001,Gou2007,Fan2008,Ando2008} is
the fact that the electron cooling is overwhelmingly dominated by the
interaction with the external supernova photons.
This enables us to avoid solving the equations for $dN_e / d\gamma_e$
self-consistently.

The same expression can be used for EIC, with a slight modification: i.e.,
$F_{\nu}^{\rm syn} \to F_{\nu}^{\rm SN} = (L_{\rm SN} / 4 \pi d^2)
\delta (\nu - \varepsilon_\gamma / h)$.
Here since the optical spectrum of a supernovae can be approximated by
a black-body spectrum, which is relatively narrow-band compared with, e.g.,
a synchrotron spectrum, we have assumed that it can be represented by
a delta function in frequency.  Then the formula further simplifies to 
\begin{equation}
F_\nu^{\rm EIC} = \frac{3\sigma_T}{64 \pi^2 R^2}
\left(\frac{h\nu}{\varepsilon_\gamma}\right)^2
\frac{L_{\rm SN}}{\nu d^2}
\int d\gamma_e \gamma_e^{-2} \frac{dN_e}{d\gamma_e}
g\left( \frac{h\nu}{4 \gamma_e^2 \varepsilon_\gamma} \right).
\end{equation}

\subsection{Results}
\label{sub:Results}

\begin{figure}
\begin{center}
\includegraphics[width=8.4cm]{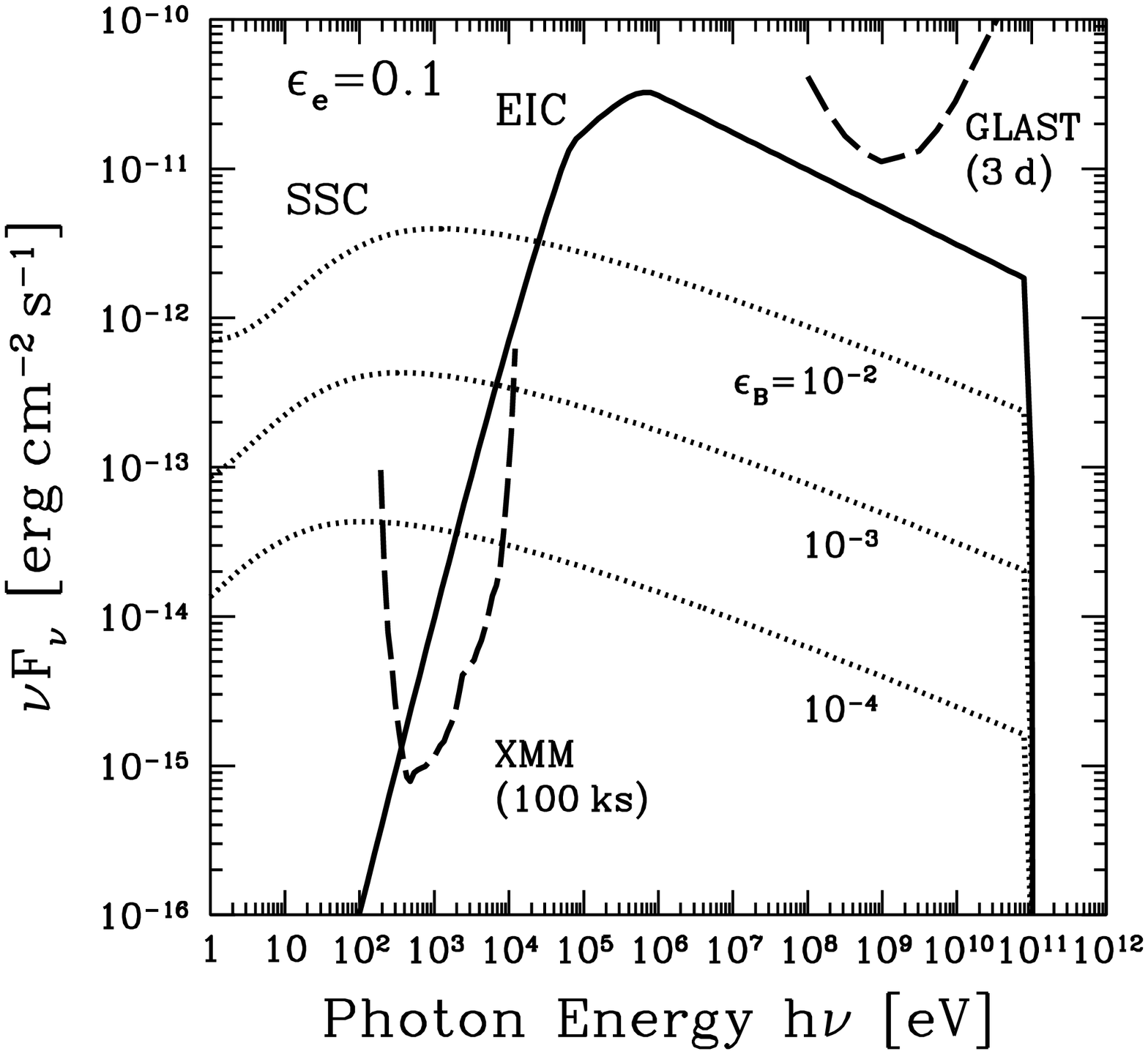}
\includegraphics[width=8.4cm]{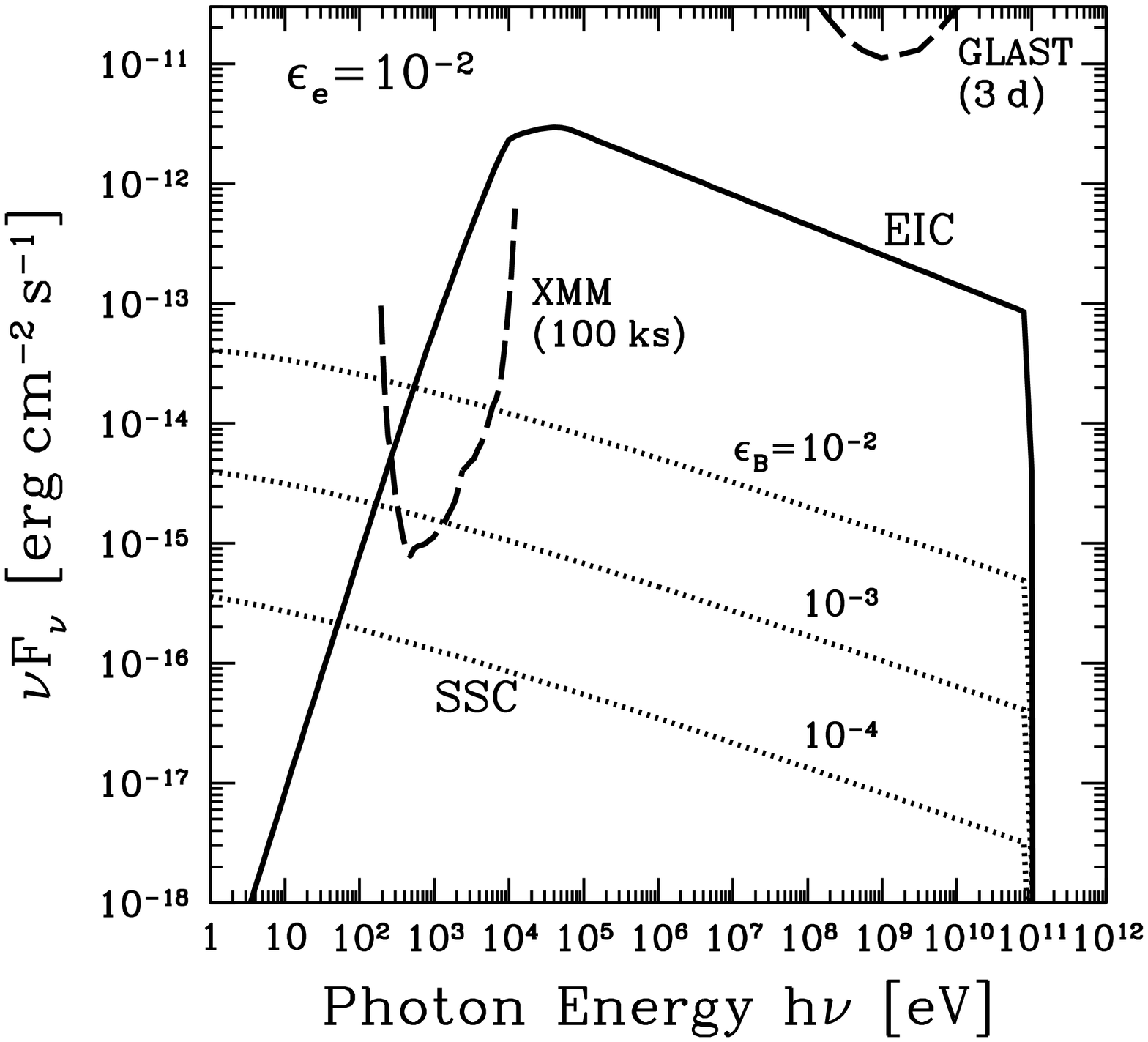}
\caption{Flux of EIC (solid) and SSC plus synchrotron (dotted) photons
 from a source at $d = 100$ Mpc, for $\epsilon_e = 0.1$ ({\rm top}),
 and $\epsilon_e = 10^{-2}$ ({\rm bottom}).  Three dotted curves in
 each panel correspond to three different values of $\epsilon_B$ as
 labeled.  The sensitivities of the {\it XMM-Newton}
 X-ray satellite as well as the {\it GLAST} gamma-ray satellite are shown
 as dashed curves.  Note that the {\it GLAST} sensitivity for a 3-d
 exposure is obtained by a simple scaling as $t^{-1/2}$ (see text).}
\label{fig:spect_e}
\end{center}
\end{figure}

In Figure~\ref{fig:spect_e}, we show the photon flux $\nu F_\nu$ 
of the electron component as a function of photon energy, for a hypernova
at a distance of $d = 100$ Mpc. 
The top and bottom panels correspond to the case of $\epsilon_e = 0.1$
(fast cooling) and $10^{-2}$ (slow cooling), respectively.
We also plot the sensitivities of the {\it XMM-Newton}
X-ray satellite (for a 100-ks exposure) and the {\it Gamma-ray Large Area
Space Telescope (GLAST)} (for a 3-d exposure); both exposure times are
comparable to $t_{\rm dyn}$.
Note that we obtained the 3-d {\it GLAST} sensitivity from the published
sensitivity of a one-year all sky survey (equivalent to a 70-d exposure to
each source),\footnote{http://www-glast.slac.stanford.edu/} by simply
rescaling as $F_{\rm sens} \propto t^{-1/2}$. This is valid strictly only
for the low-energy region where the sensitivity is limited by background.
For the high-energy range, which is limited by photon counts, the 
sensitivity scales as $F_{\rm sens} \propto t^{-1}$ instead. However,
understanding at which energy the transition happens requires a more
careful study of the detector performance. For simplicity, we used 
the $t^{-1/2}$ dependence in the figure, because the detection would 
be dominated by low-energy photons around 100 MeV.

\subsubsection{External inverse-Compton scattering}

\begin{figure}
\begin{center}
\includegraphics[width=8.4cm]{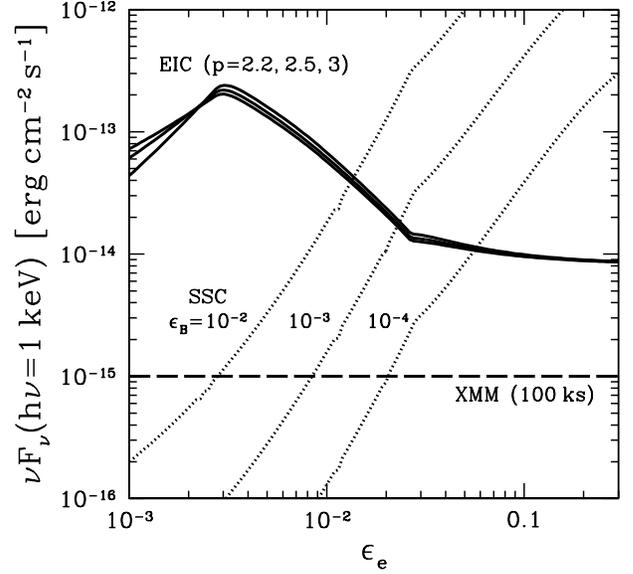}
\caption{Flux of EIC (solid) and SSC (dotted) photons at 1 keV as a
  function of $\epsilon_e$ for $d = 100$ Mpc.  Three EIC (SSC) curves
  correspond to different values of  $p$ ($\epsilon_B$), as labeled.
  The sensitivity of the current X-ray satellites are also shown as a
  horizontal line.}
\label{fig:F1keV}
\end{center}
\end{figure}

The EIC flux has a very characteristic shape in the X-ray range, where
its spectrum is quite hard.
This is because the typical inverse-Compton frequency corresponding to
$\gamma_m$ and $\gamma_c$ are above the X-ray region---i.e.,
$h \nu_m^{\rm EIC} \approx 4\Gamma \gamma_m^2 \varepsilon_\gamma / 3 =
350 (\epsilon_e/0.1)^2$ keV and $h \nu_c^{\rm EIC} \approx 4 \Gamma
\gamma_c^2 \varepsilon_\gamma / 3 = 25$ keV---and 
in the X-ray range we see the low-energy tail of the external
photons scattered by the least energetic electrons.
This is consistent with the considerations of \citet{Waxman1999}, 
while the authors focused on emission from thermal electrons.
In addition, as the electron distribution is determined by cooling by
the EIC process with seed photons provided by an external source, the
EIC flux is largely independent of $\epsilon_B$, which is another useful
feature enabling us to probe $\epsilon_e$ in a robust manner with
X-ray measurements.
In Figure~\ref{fig:F1keV}, we show $\nu F_\nu^{\rm EIC}$ at 1 keV
(solid curves) as a function of $\epsilon_e$, for $d = 100$ Mpc.
The shape of $\nu F_\nu^{\rm EIC}$ is roughly divided into the
following three different regions.
\begin{enumerate}
\item 
The region $\epsilon_e < 3 \times 10^{-3}$ corresponds to electrons
in the slow-cooling regime and $h \nu_m < 1 ~ \mathrm{keV} < h \nu_c$.
Here the typical Lorentz factor of the electrons that emit EIC photons
at 1 keV ($\gamma_{\rm keV}$) satisfies $\gamma_m < \gamma_{\rm keV} <
\gamma_c$, for which the electron distribution is given by the first
expression of equation~(\ref{eq:Ne slow cooling}).
From this expression, it is clear that increasing $\epsilon_e$ (or
equivalently $\gamma_m$) makes the 1 keV flux larger, and the flux 
depends, albeit weakly, on the electron spectral index $p$.
\item
The middle region $3 \times 10^{-3} < \epsilon_e < 0.03$
corresponds to $1 ~ {\rm keV} < h \nu_m < h \nu_c$ (slow cooling).
In this case the flux is dominated by the tail emission from electrons
with $\gamma_m$.  Since the spectrum is hard at 1 keV ($\nu F_{\nu} 
\propto \nu^2$), it rapidly decreases as $\nu_m$ goes away from 1 keV 
(with increasing $\epsilon_e$).
\item
In the region with $\epsilon_e > 0.03$, the flux is almost independent
of $\epsilon_e$.  The electrons are in the fast cooling regime, 
$1 ~ {\rm keV} < h \nu_c < h \nu_m$.
In this case, the flux at 1 keV is mostly due to tail emission of
electrons with $\gamma_c$, whose numbers are little dependent on
$\gamma_m$, as clearly seen in the first expression of
equation~(\ref{eq:Ne fast cooling}).
\end{enumerate}
In all cases, the soft X-ray flux is fairly high and robust, and would
be an excellent target for {\it XMM-Newton}.

\begin{figure}
\begin{center}
\includegraphics[width=8.4cm]{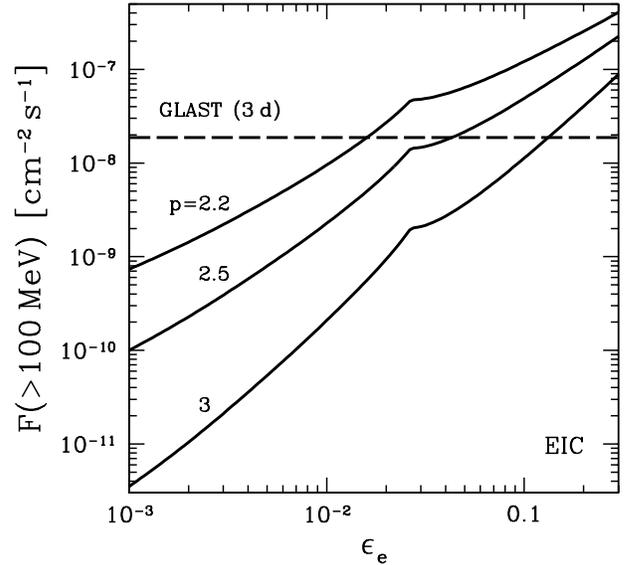}
\caption{Flux of EIC photons integrated above 100 MeV as a function of
 $\epsilon_e$ for various values of $p$ as labeled, for $d = 100$ Mpc.
 The {\it GLAST} sensitivity for a 3-d exposure is shown as a horizontal
 line.}
\label{fig:F100MeV}
\end{center}
\end{figure}

In the GeV gamma-ray range, the EIC flux may be detected by {\it
GLAST} if $\epsilon_e$ is large enough (Figure~\ref{fig:spect_e}).
Although we have not assumed here any cutoff of the electron spectrum, 
the presence of such a cutoff as well as the spectral shape
(characterized by $p$) could be tested with {\it GLAST}.
The sharp cutoff seen around $10^2$ GeV in Figure~\ref{fig:spect_e} is
due to absorption by supernova photons, leading to 
electron-positron pairs (see discussion below).
In Figure~\ref{fig:F100MeV}, we show the EIC flux integrated above
100 MeV (for $d = 100$ Mpc) plotted as a function of $\epsilon_e$ for
three different values of $p$.
Here the {\it GLAST} sensitivity is rescaled using the $t^{-1/2}$
dependence and $t = t_{\rm dyn} \approx 3$ d, since the flux
sensitivity above 100 MeV is in the background-limited regime
(it becomes better if we use the {\it integrated} flux rather
than the {\it differential} one shown in Fig.~\ref{fig:spect_e}).
For a hard electron spectrum, the EIC flux could be large in the
GeV energy region for reasonable values of $\epsilon_e$, so that {\it
GLAST} should be able to probe the spectrum, including the maximum
acceleration energy of the electrons.

\subsubsection{Synchrotron self-Compton scattering}

The SSC flux depends sensitively on both $\epsilon_e$ and $\epsilon_B$,
as shown in Figures~\ref{fig:spect_e} and \ref{fig:F1keV}. This would be a 
dominant component in the X-ray range, well within the sensitivity limit
of X-ray telescopes, if both $\epsilon_e$ and $\epsilon_B$ are large enough.
Since the spectral shape is very different from that of the EIC component, 
these components of two different origins should be easily distinguishable.
In the GeV regime, on the other hand, the SSC flux is likely to be
subdominant, unless $\epsilon_B$ is comparable to $\epsilon_e$ and as
large as $\sim$0.1.

As briefly argued at the end of \S~\ref{sub:Synchrotron radiation}, if
the generated magnetic fields were highly inhomogeneous on very small
scales (although we believe this is unlikely in the current setup),
then the synchrotron radiation would be replaced by another radiation
component---jitter radiation \citep{Medvedev2000}.
The spectrum of the inverse-Compton photons off the jitter radiation
would be similar to the jitter spectrum, which results in a harder
spectrum than SSC at low frequency.
Therefore, the difference ($F_\nu \propto \nu$ vs $\nu^{1/3}$) might
appear in the X-ray regime and could be tested.

\subsubsection{Regenerated emission due to pair production}
\label{subsub:Regenerated emission due to pair production}

For photon energies $\gtrsim 10^2$ GeV, the $\gamma\gamma \to e^+e^-$
opacity against surrounding supernova photons becomes greater than 
unity \citep[e.g.,][]{Stecker1992}.  Thus, these high-energy photons 
cannot escape from the source, giving a sharp cutoff in the spectrum 
as shown in Figure~\ref{fig:spect_e}. The resulting $e^\pm$ pairs, 
having Lorentz factor above $\sim$10$^5$, then radiate synchrotron 
photons over a wide frequency range.  The inverse-Compton scattering 
of supernova photons by these pairs occurs mainly in the Klein-Nishina 
regime, where the scattering cross section is suppressed 
\citep{Rybicki1979}, and can be neglected here.

\begin{figure}
\begin{center}
\includegraphics[width=8.4cm]{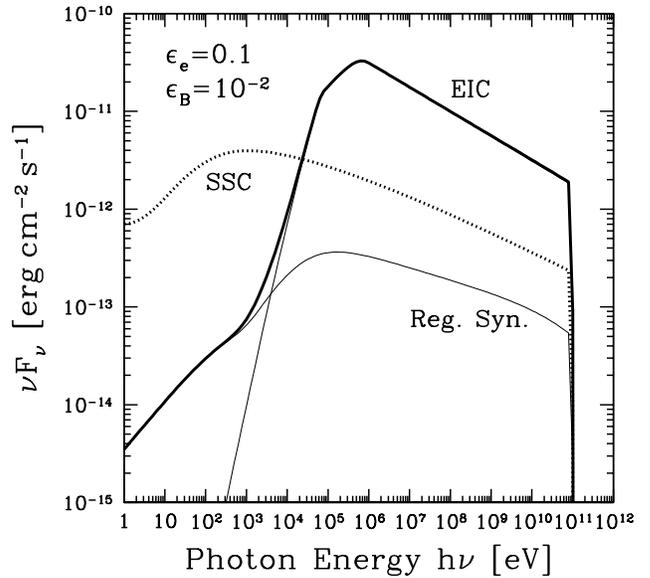}
\caption{Synchrotron spectrum from electron-positron pairs (``regenerated''
emission) produced by the interaction between high-energy EIC photons 
($h \nu > 10^2$ GeV) and supernova photons (labeled as ``Reg. Syn.'').  
This flux is compared with other radiation components, SSC and EIC, for 
the fiducial parameter set and $d = 100$ Mpc.}
\label{fig:spect_reg}
\end{center}
\end{figure}

The resulting pairs rapidly lose their energy through radiation, and
thus they are in the strongly fast-cooling regime.  We computed the
synchrotron spectrum from these pairs, and show the results
in Figure~\ref{fig:spect_reg} for the fiducial case and $d = 100$ Mpc.
We find that this radiation component is significantly lower than both
the SSC and EIC components for all the available energy ranges.
While we show here only the result for the fiducial case, the same  
conclusion would apply to other parameter sets, because the dependence 
of the regenerated emission on $\epsilon_e$ or $\epsilon_B$ is more or 
less the same as that of the SSC or EIC emission.  In addition, if the 
EIC spectrum had any cutoff above $10^2$ GeV, this would reduce even
further the regenerated flux.  Thus, the regeneration emission component 
can be safely ignored in the present discussion.

\section{Radiation from secondary electrons and proton acceleration}
\label{sec:Radiation from secondary electrons and proton acceleration}

Recently, \citet{Wang2007} and \citet{Budnik2008} suggested that hypernovae 
could account for some of the cosmic-ray flux that we measure today, and also
investigated the possibility of the associated neutrino emission and 
detection in order to further test such a scenario.  \citet{Asano2008} 
studied another consequence of this mechanism, the gamma-ray emission from 
secondary electrons produced by interactions involving accelerated protons.
Their conclusion is that the synchrotron flux from these secondary
electrons could be detected by modern X-ray telescopes.
Here we further extend their study, comparing the results with that of
the primary electron component, and investigate which emission components 
would dominate in different energy ranges, depending on the parameters.
As the origin of the secondary electrons, we consider decays of charged
pions produced by interactions between protons and supernova photons,
e.g., $p \gamma \to n \pi^+$ (labeled simply as $p\gamma$), and pair
production through the Bethe-Heitler process, $p \gamma \to p e^+
e^-$ (labeled as BH).  The BH process was not taken into account in the 
computation of \citet{Asano2008}.

Following \citet{Wang2007} and \citet{Asano2008}, we assume that the
total energy of cosmic-ray protons is $E_{\rm CR} \simeq  E_K / 6$, and
the energy distribution follows a $\varepsilon_p^{-2}$ spectrum.
The maximum acceleration energy of the protons is assumed to be given
by $\varepsilon_{p,{\rm max}} = e B R \beta \simeq 2 \times 10^{18}
(\epsilon_B / 10^{-2})^{1/2} ~ {\rm eV}$.
We use again $\epsilon_B = 10^{-2}$ as our fiducial value, to compare
with our primary electron radiation. Note that \citet{Asano2008} used 
$\epsilon_B = 0.1$. Due to this our estimate results in slightly 
smaller fluxes than those of \citet{Asano2008}; as we argue below, 
the main dependence on $\epsilon_B$ is expected to be weak.
The threshold energy for pion production through $\Delta$-resonance is
roughly given by $\varepsilon_{p,{\rm th}}^{p\gamma} = 0.3 ~ {\rm
GeV}^2 / \varepsilon_\gamma \simeq 10^{17}$ eV; thus protons with
energies near $\varepsilon_{p,{\rm max}}$ can produce charged pions
through this mechanism.
The energy-loss time scale of the protons is given by $t_{p\gamma} =
(0.2 \sigma_{p\gamma} n_\gamma c)^{-1} \simeq 10^6$ s, where
$\sigma_{p \gamma} = 5 \times 10^{-28}$ cm$^2$ (almost independent
of energy) is the $p\gamma$ cross section and the factor 0.2
represents the energy lost by the proton by each interaction.
The fraction of the total proton energy $E_{\rm CR}$ taken away by
this process may then be estimated by $f_{p\gamma} = t_{\rm dyn} /
t_{p\gamma} \approx 0.2$.
We assume that half of the outcome of $p\gamma$ interaction is in
the form of charged pions, and that they carry off 20\% of the parent 
proton energy, conserving the original $\epsilon_\pi^{-2}$ spectrum.

\begin{figure}
\begin{center}
\includegraphics[width=8.4cm]{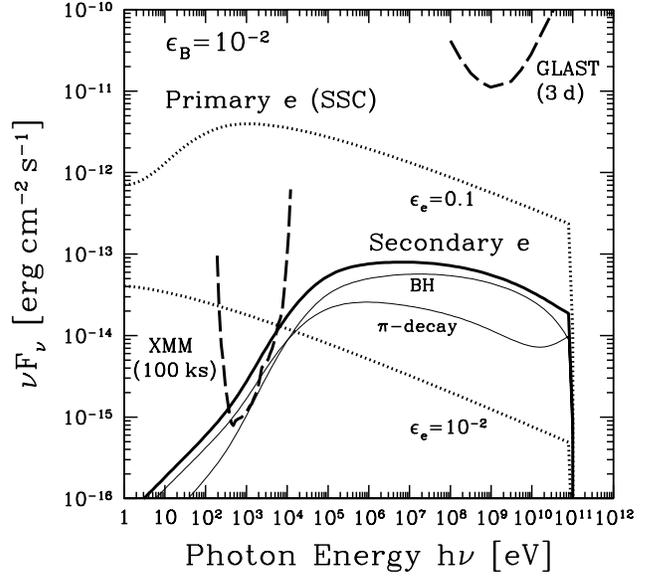}
\caption{Flux of synchrotron photons from secondary electrons produced
by proton interactions (solid), for $\epsilon_B = 10^{-2}$ and $d =
100$ Mpc.  Components from $\pi$ decays and the Bethe-Heitler (BH) 
process are shown as thin solid curves.  For comparison, the SSC spectrum 
from primary electrons for $\epsilon_e = 0.1$ and $10^{-2}$ are also 
shown as dotted curves.}
\label{fig:spect_p}
\end{center}
\end{figure}

Secondary electrons or positrons are produced by pion decays and we
assume that their energies are 1/4 of that of the parent pions 
($\varepsilon_e = 0.25 \varepsilon_\pi = 0.05 \varepsilon_p$), since
a charged pion eventually decays into four light particles (one 
charged lepton and three neutrinos).  The electron spectrum is again 
approximately represented by a power law with the same index, i.e., 
$\varepsilon_e^{-2}$ for $6 \times 10^{15} < \varepsilon_e / {\rm eV} 
< 10^{17} (\epsilon_B / 10^{-2})^{1/2}$.
These electrons rapidly lose their energy radiatively, and we obtain
the final energy distribution by equating radiative energy loss and
injection due to $\pi$ production and decay.
The typical energy of the synchrotron photons corresponding to
$\varepsilon_e = 10^{16}$ eV is $\nu_{\rm syn} \sim 10^{13}$ eV.
Thus, the synchrotron spectrum from these secondary electrons is very
hard and peaks in the multi-TeV region, where absorption by supernova
photons is significant. A third generation of electrons and positrons 
is then produced, and they again radiate synchrotron photons, as discussed 
in \S~\ref{subsub:Regenerated emission due to pair production}.
We have included this process as well, and show the spectrum due to this
$p\gamma$ origin in Figure~\ref{fig:spect_p} (labeled as ``$\pi$-decay'').
The SSC spectra due to primary electrons  are also shown for comparison, 
for two cases of $\epsilon_e = 0.1$ and $\epsilon_e = 10^{-2}$.
Although the flux itself from proton-originated leptons could be
larger than the sensitivity of {\it XMM-Newton} for these parameters, a
result consistent with that of \citet{Asano2008}, it is
much less significant than the SSC component from primary electrons.

The Bethe-Heitler (BH) process can also be important for producing secondary
electrons which radiate synchrotron photons in the relevant energy range.
The BH cross section is $\sigma_{\rm BH} = (28/9) \alpha r_e^2 \ln
[2 \varepsilon_p \varepsilon_\gamma / m_p m_e c^4 - 106/9 ]$, where
$\alpha$ is the fine-structure constant and $r_e = e^2 / m_e c^2$ is
the classical electron radius.
The threshold energy of this interaction is then $\varepsilon_{p,{\rm
th}}^{\rm BH} = (115/18) m_e m_p c^4 / \varepsilon_\gamma \simeq
10^{15}$ eV, much smaller than the $p\gamma$ threshold
$\varepsilon_{p,{\rm th}}^{\rm p\gamma} \simeq 10^{17}$ eV.
At each BH interaction, the proton loses energy by producing
electron-positron pairs, and the magnitude is $\Delta \varepsilon_p =
2 m_e c^2 \gamma_{\rm cm}$, where $\gamma_{\rm cm} = (\varepsilon_p +
\varepsilon_\gamma) / ( m_p^2 c^4 + 2 \varepsilon_p \varepsilon_\gamma
)^{1/2} \approx \varepsilon_p / m_p c^2$ for the relevant $\varepsilon_p$ 
range.  Correspondingly, the energy of the produced electron or positron 
is $\varepsilon_e = m_e c^2 \gamma_{\rm cm} \approx \varepsilon_p m_e /
m_p = 5 \times 10^{-4} \varepsilon_p$.  The energy-loss time scale of 
the proton is then given by $t_{\rm BH} = ( \sigma_{\rm BH} n_\gamma c 
\Delta \varepsilon_p / \varepsilon_p )^{-1}$.
The fraction of the total proton energy $E_{\rm CR}$ lost by this
process is estimated as $f_{\rm BH} = t_{\rm dyn} / t_{\rm BH}$, which
is typically $\sim$0.01--0.03.
We obtain the injection spectrum of these electrons and positrons by
multiplying the proton spectrum by $f_{\rm BH}$.  Since $f_{\rm BH}$ 
is almost constant, this injection spectrum is again close to 
$\varepsilon_e^{-2}$ for $5 \times 10^{11} < \varepsilon_e / {\rm eV} 
< 10^{15} (\epsilon_B / 10^{-2})^{1/2}$; the final spectrum
of the pairs is then obtained as the steady state solution of
a diffusion equation including radiative energy losses.
We find that the regenerated emission is negligible because the
typical electron energy is much smaller than the $p\gamma$ case.
We show the synchrotron spectrum from the BH process in 
Figure~\ref{fig:spect_p}, and find that it is indeed comparable to
$p\gamma \to n\pi^+$ component.
The lower fraction of total energy carried away ($f_{\rm BH} \ll
f_{p\gamma}$) is compensated by the fact that pairs have lower
energies, which radiate many photons in the relevant energy range such
as soft X-rays.

Compared to the SSC photon component from the primary electron population, 
the photon component from a proton origin is significantly less important.
This is also the case for other values of $\epsilon_B$ because both
components are related to synchrotron radiation, and the relative 
importance would not change much. Considering values of $\epsilon_e$
much smaller than $10^{-2}$ would change the proton to SSC ratio in
favor of the former, but even in this case, the EIC component, which 
is not plotted explicitly in Figure~\ref{fig:spect_p}, would still be
very large in the soft X-ray regime as shown in Figure~\ref{fig:F1keV}.
This remains the case as long as $\epsilon_e \gtrsim 10^{-3}$.
Using steeper spectrum or smaller maximum acceleration energy for the
protons simply reduces the proton yields;
note that especially the $p\gamma$ process is very sensitive to these
values.
In the calculation above, we used a very hard proton spectrum
$\varepsilon_p^{-2}$ and maximum efficiency for the acceleration,
which is the best case for the proton component.
Thus, the goal of testing proton acceleration in hypernovae and its
connection to cosmic rays with X-ray or gamma-ray telescopes is hard
to achieve, since the photon signatures would be largely swamped by a
strong primary electron component.  Neutrinos would remain the most
reliable means of testing such a scenario.

Although difficult, we also mention two possibilities that might
enable detection of a proton component in photon radiation.
First, if one had situations such that $\epsilon_e \ll 10^{-3}$, where
the bulk of the primary electrons might be nonrelativistic, whereas
protons were accelerated efficiently in sufficiently large magnetic
fields, then it might be possible to suppress the primary component
compared with the secondary.
Second, since the proton cooling times are longer than that of
electrons, late time observations might help detection of a proton
component, although adiabatic cooling and decaying flux levels would
require long-time integrations.

\section{Conclusions}
\label{sec:Conclusions}

We have studied the radiation from the mildly relativistic ejecta 
associated with hypernovae. The energy associated with this portion
of the ejecta, about $10^{50}$ erg, starts to dissipate at radii of 
about 10$^{16}$ cm, where electrons and protons are shock accelerated.
The electrons radiate photons via synchrotron, synchrotron-self
Compton (SSC) of synchrotron photons, and external inverse-Compton 
(EIC) scattering of hypernova photons. The radiation is most prominent
in the X-ray and GeV ranges, detectable for sources at $d < 100$ Mpc.
We also studied the radiation from second and third generations of
electrons and positrons that are associated with accelerated protons.
The interactions that produce these leptons are the photopion and 
Bethe-Heitler (BH) processes, and both of these would give fluxes above
the sensitivity limit of modern X-ray telescopes.
However, we find that both of these components are hidden by a large
flux due to the primary electron population mentioned above, for most
realistic combinations of relevant parameters.  Values of $\epsilon_e
\ll 10^{-3}$ or very long lasting observations may improve the detection
prospects of this component.

The most promising energy range for detecting the electron SSC and EIC 
emission components is in the soft X-ray range.  If both $\epsilon_e$ 
and $\epsilon_B$ are reasonably large, the SSC component dominates, 
while otherwise the EIC does (Fig.~\ref{fig:F1keV}). The regenerated 
emission component from electron-positron pairs produced by $\gamma\gamma$
absorption of $\gtrsim 10^2$ GeV photons is found to be always negligible.
A robust feature of the EIC emission is that it provides fairly large flux 
in the X-ray regime, for most combinations of the parameters $\epsilon_e$, 
$\epsilon_B$ and $p$, if the supernova occurs at distances $d \leq 100$ Mpc.
When the two components are comparable, it may be easy to distinguish
them through their very different spectral shapes. The GeV flux is 
dominated by the EIC component. Observations with {\it GLAST} would make
it possible to measure $p$ independently, if $\epsilon_e$ is large enough, 
and wold also be able to probe for the  spectral cutoff corresponding to 
the maximum acceleration energy of the radiating electrons.

Finally, we briefly comment on the dependence of the results on the
progenitor wind mass-loss rate.  This is interesting in particular 
because \citet{Campana2006} estimated a substantial mass-loss rate 
$\dot M \sim 10^{-4} M_{\sun}$ for the progenitor of GRB 060218/SN 2006aj, 
which is an order of magnitude larger than the nominal value adopted here.
Such a higher mass loss rate decreases the dissipation radius $R$ and the
dynamical time scale $t_{\rm dyn}$ by an order of magnitude. Even if
we decrease the supernova luminosity to $L_{\rm SN} = 10^{42}$ erg s$^{-1}$,
an order of magnitude below the nominal value adopted above, while keeping 
the other parameters the same, we find that both the SSC and EIC photon 
fluxes increase by about 2--3 orders of magnitude. This more than offsets
the decrease by a factor 3 of the sensitivities of X-ray telescopes caused 
by the smaller $t_{\rm dyn}$, implying a very significant further improvement 
in the prospects for detection.

%%%%%%%%%%%%%%%%%%%%%%%%%%%%%%%%%%%%%%%%%%%%%%%%%%%%%%%%%%%%%%%%%%%%
\acknowledgments

We thank Katsuaki Asano for valuable comments.
This work was supported by the Sherman Fairchild Foundation (SA) and
by NSF AST0307376 and NASA NAG5-13286 (PM).

%\appendix

\end{document}